%% file: sample-authordraft.tex
\documentclass[sigconf, authorversion]{acmart}

\usepackage{balance}
\usepackage{placeins}
\usepackage{enumitem}
\usepackage{microtype}

\usepackage{bm}
\usepackage{array}
\usepackage{booktabs}
\usepackage{multirow}
\usepackage{makecell}
\usepackage{adjustbox}

\usepackage{listings}
\usepackage{xcolor,colortbl}
\definecolor{tableheadercolour}{HTML}{F7E6CF}

\definecolor{codegreen}{rgb}{0,0.6,0}
\definecolor{codegray}{rgb}{0.5,0.5,0.5}
\definecolor{codepurple}{rgb}{0.58,0,0.82}
\definecolor{backcolour}{rgb}{0.95,0.95,0.92}

\lstdefinestyle{mystyle}{
    backgroundcolor=\color{backcolour},   
    commentstyle=\color{codegreen},
    keywordstyle=\color{magenta},
    numberstyle=\tiny\color{codegray},
    stringstyle=\color{codepurple},
    basicstyle=\ttfamily\footnotesize,
    breakatwhitespace=false,         
    breaklines=true,                 
    captionpos=b,                    
    keepspaces=true,                 
    numbers=left,                    
    numbersep=5pt,                  
    showspaces=false,                
    showstringspaces=false,
    showtabs=false,                  
    tabsize=2
}

\newlist{researchquestions}{enumerate}{1}
\setlist[researchquestions]{label*=\textbf{RQ\arabic*.}}

\newlist{openquestions}{enumerate}{1}
\setlist[openquestions]{label*=\textbf{PSQ\arabic*.}}

\AtBeginDocument{%
  \providecommand\BibTeX{{%
    \normalfont B\kern-0.5em{\scshape i\kern-0.25em b}\kern-0.8em\TeX}}}

\copyrightyear{2025}
\acmYear{2025}
\setcopyright{acmlicensed}\acmConference[SIGCSE TS 2025]{Proceedings of the 56th ACM Technical Symposium on Computer Science Education V. 1}{February 26-March 1, 2025}{Pittsburgh, PA, USA}
\acmBooktitle{Proceedings of the 56th ACM Technical Symposium on Computer Science Education V. 1 (SIGCSE TS 2025), February 26-March 1, 2025, Pittsburgh, PA, USA}
\acmDOI{10.1145/3641554.3701911}
\acmISBN{979-8-4007-0531-1/25/02}

\begin{document}


\title[Productive Failure as a Design Paradigm Introductory Python Programming]{Investigating the Use of Productive Failure as a Design Paradigm for Learning Introductory Python Programming}

\author{Hussel Suriyaarachchi}
\orcid{0000-0002-8026-2523}
\affiliation{
\institution{Augmented Human Lab \\ National University of Singapore}
  \country{Singapore}
}
\email{hussel@ahlab.org}

\author{Paul Denny}
\orcid{0000-0002-5150-9806}
\affiliation{
  \institution{School of Computer Science \\ University of Auckland}
  \city{Auckland}
  \country{New Zealand}
}
\email{paul@cs.auckland.ac.nz}

\author{Suranga Nanayakkara}
\orcid{0000-0001-7441-5493}
\affiliation{
\institution{Augmented Human Lab \\National University of Singapore}
  \country{Singapore}
}
\email{suranga@ahlab.org}

\renewcommand{\shortauthors}{Hussel Suriyaarachchi, Paul Denny, \& Suranga Nanayakkara}


\begin{abstract}
Productive Failure (PF) is a learning approach where students initially tackle novel problems targeting concepts they have not yet learned, followed by a consolidation phase where these concepts are taught. Recent application in STEM disciplines suggests that PF can help learners develop more robust conceptual knowledge. However, empirical validation of PF for programming education remains under-explored. In this paper, we investigate the use of PF to teach Python lists to undergraduate students with limited prior programming experience. We designed a novel PF-based learning activity that incorporated the unobtrusive collection of real-time heart-rate data from consumer-grade wearable sensors. This sensor data was used both to make the learning activity engaging and to infer cognitive load. We evaluated our approach with 20 participants, half of whom were taught Python concepts using Direct Instruction (DI), and the other half with PF. We found that although there was no difference in initial learning outcomes between the groups, students who followed the PF approach showed better knowledge retention and performance on delayed but similar tasks. In addition, physiological measurements indicated that these students also exhibited a larger decrease in cognitive load during their tasks after instruction. Our findings suggest that PF-based approaches may lead to more robust learning, and that future work should investigate similar activities at scale across a range of concepts.
\end{abstract}



\maketitle

\input{paper-body}

\balance

\section*{Acknowledgements}
This work was supported by the Assistive Augmentation research
grant under the Entrepreneurial Universities (EU) initiative of New Zealand and the Tier 1 Ministry of Education (MOE) Academic Research Fund (AcRF).

\bibliographystyle{ACM-Reference-Format}
\bibliography{sample-base}

\end{document}

%% file: paper-body.tex
\section{Introduction} \label{ch:1}
Productive Failure (PF) is an instructional approach where students initially engage with problems targeting concepts they have not yet learned, followed by a phase of consolidation where these concepts are taught \cite{kapur2008productive, kapur2012designing}. The rationale behind PF is that initial struggle and failure can activate prior knowledge and create a fertile ground for subsequent learning, leading to deeper understanding and better retention of concepts \cite{kapur2016examining}.  

Since its introduction, PF has gained recognition among researchers and educators for its success in STEM disciplines, most popularly mathematics and physics, where it has been shown to promote more robust learning outcomes compared to traditional instructional methods~\cite{kapur2012designing, sinha2021problem, trueman2014productive}.  Despite these successes in other fields, there has been little empirical work exploring the application of PF in computing or programming education in comparison to more traditional Direct Instruction (DI) approaches.
Recent work indicates that PF may be a promising approach for computing education~\cite{savelson2023how, thorgeirsson2022does}, but there is a need for more work on designing appropriate tasks and for objective data on its effects.

To address this gap, in this paper, we describe our design and development of a novel PF-based activity targeting lists in Python. This activity was informed by an iterative, student-centred design approach involving pilot studies with participants of varied programming proficiencies.  We incorporate real-time data collection during the learning activity, utilising consumer-grade wearable sensors to measure physiological data.  This data was used by students as part of the programming task, and it also allowed us to infer cognitive load.

We conducted a controlled study involving 20 participants to evaluate the impact of our PF-based activity on learning outcomes and cognitive load, in comparison to traditional Direct Instruction (DI), measuring performance on a two-week delayed post-test task. Our research is guided by the following questions:
    
\begin{researchquestions}
    \item  How does the choice of pedagogical design impact initial task performance and subsequent success on a delayed task?
    \item To what extent do physiological measures of cognitive load change after students receive programming instruction, and does PF influence these changes?
    \item What are students' preferences and perceptions regarding PF and DI?
\end{researchquestions}


Our findings show that students performed similarly on programming tasks immediately after instruction, regardless of the pedagogical strategy. However, those using the PF approach performed better on the same tasks two weeks later. We also explore the implications for reducing cognitive load and its link to improved performance.

\section{Related work}
Knowledge of computing, and specifically programming, has become a core skill that shapes how students think and tackle problems across various subject domains \cite{farah2020bringing}. There has been considerable research interest in exploring and developing effective pedagogical approaches for teaching programming \cite{denny2019research}.

\subsection{Productive Failure in Computing Education}

Productive Failure (PF) \cite{kapur2008productive} is an established instructional framework that lends itself to constructionist learning \cite{harel1991constructionism} in which students engage in a Problem-Solving followed-by-instruction (PS-I) \cite{loibl2017towards, schalk2018problem} learning design to construct new knowledge. In recent years, PF has begun to appear in computing education, with emerging use in introductory CS1/CS2 undergraduate programs. 

Preliminary work by Thorgeirsson et al. \cite{thorgeirsson2022does} explored the use of PF to introduce programming concepts such as sorting algorithms, graph theory, and cluster assignment. Their failure-driven learning experience was facilitated through prompts that deliberately nudged students towards suboptimal solutions within a visual programming environment named Algot \cite{thorgeirsson2021algot}. Similarly, Savelson and Muldner \cite{savelson2020student} delivered their PF strategy on sorting using a custom interface of the popular block-based programming platform Snap!\footnote{\url{https://snap.berkeley.edu/}}. These studies showed favourable effects in improving students' constructive reasoning and curiosity compared to DI approaches while maintaining similar frustration levels.

Although this recent work indicated positive learning behaviours, there was inconclusive evidence on the impact of PF on learner performance and conceptual understanding. Steinhorst et al. \cite{steinhorst2023exploring, steinhorst2024recognizing} also noted these concerns in their PF interventions on computing education. Their results indicated that the learning advantages of PF over DI observed in other fields, with respect to conceptual knowledge and task performance, could not be replicated for computing concepts on operating systems. Thus, considering the infancy of PF in computing education and its known potential for robust learning, there is a need for more empirical work to reliably determine its efficacy.

\subsection{Understanding Learning}
Understanding what it means to learn is a particularly critical but challenging task in applied educational settings \cite{berliner2002comment}. The assessment of learning and its outcomes is typically informed by the achievement of curricular goals and skills \cite{sridhar2020progression}. Instructors rely on diagnostic and summative strategies such as self-reported measures and standardised tests to gauge learning performance \cite{pekrun2014self, popham2001truth}. While these conventional academic metrics may be functional, they fail to accurately represent learning processes \cite{resnick1992assessing}. Thus, researchers need to confront the task of providing more reliable and objective measures of learning \cite{augereau2019experimental}.

With learning being a complex and multi-faceted process embodying both the body and mind, there has been much interest in exploring its relationship to various cognitive and affective states \cite{dahlstrom2019showing, sridhar2019going, tan2021case}. Variations observed in students' physiology are increasingly recognised as effective indicators of factors that underpin learning, such as cognitive load \cite{kaewkamnerdpong2016framework, mills2017put}. For example, heart rate variability (HRV) is one such physiological measure that is closely linked to cognitive load \cite{eija2010}. Prior work exploring cognitive load in computing and programming tasks found that experiencing a reduced cognitive load tends to improve student learning \cite{sands2019addressing, sweller1994cognitive}. 

In light of the recent wave of consumer-friendly physiological sensors and smart wearables, the possibility of seamlessly monitoring cognitive and physiological functions in the classroom lies within reach. However, introducing  tools that 
are not directly aligned with learning activities may be distracting for students and 
disrupt authentic learning processes \cite{suriyaarachchi2022primary}. 

In this work, we explore the use of Productive Failure in computing education and investigate its potential in an introductory Python undergraduate course. We incorporate wearable consumer sensors 
as essential parts of  learning activities to make their presence intuitive to students and to enable the unobtrusive collection of physiological data on authentic learning processes.

\section{PF Design \& Development}
To explore our research questions, we developed a Python programming task
focused on lists.
The choice of concept was motivated by ACM's curriculum guidelines for undergraduate computer science education \cite{acm2013}, which identify lists as a popular CS1 construct.   Lists are also an ideal topic for designing problem-solving contexts that embody Kapur et al.'s recommendations for PF learning experiences, such as accommodating various solution approaches \cite{kapur2012designing}.

\subsection{Programming Task} \label{ch:3.1}
For our study, we created an open-ended challenge that targeted Python lists and involved storing data from a heart-rate sensor. 

\subsubsection{Python List Task}
The task on lists was developed as a simplified sliding-window problem on a live stream of heart-rate data and is accessible online\footnote{\url{https://bit.ly/pf-list-task}}. 
Students are required to explore using their existing knowledge of data structures and variables to maintain a record of the ten most recent heart rate readings and ensure that this data stored is always up-to-date. While the canonical solution involves using a list to `append' and `pop' data based on its length, our task design also supports the generation of suboptimal representations and solution methods using constructs such as strings and temporary variables.

\begin{figure}[b]
    \centering
    \vspace{-5pt}
    \includegraphics[width=\linewidth]{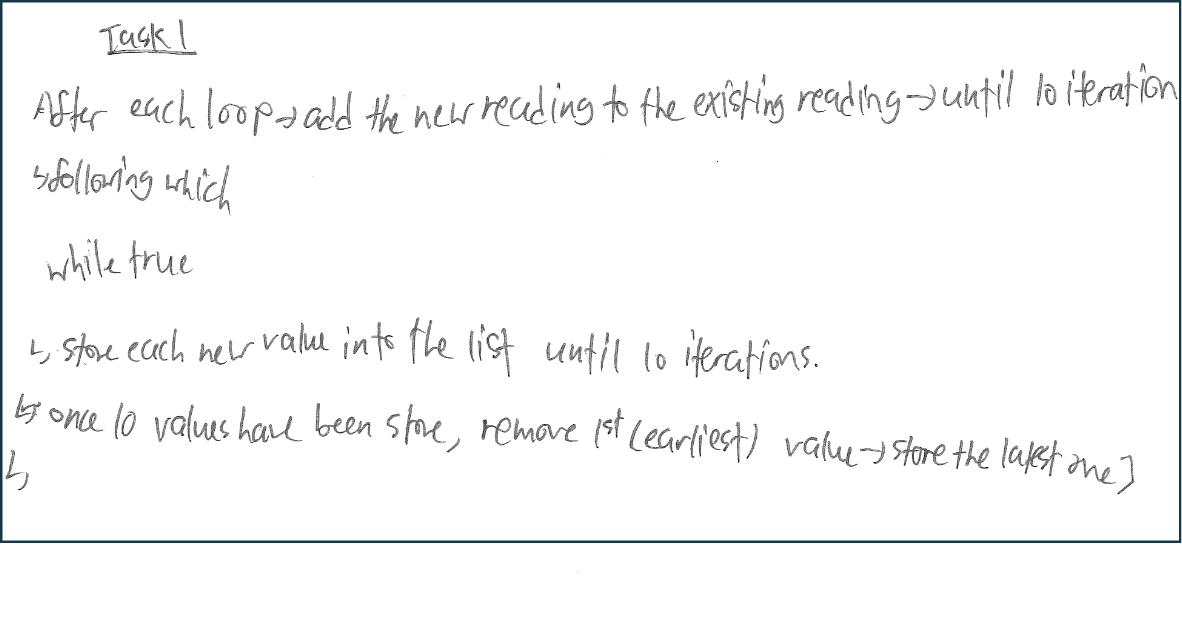}
    \vspace{-28pt}
    \caption{A handwritten solution collected during the pilot evaluation illustrating exploration of the problem.}    
    \label{fig:student_rsm}
    \vspace{-10pt}
\end{figure}

\subsubsection{Pilot Evaluation} 
After initial development of the task, and prior to our main evaluation, we conducted a pilot study to ensure the task was reliably understood and successfully promoted problem-solving behaviours representative of a PF learning exercise. This pilot was structured as a think-aloud study with 11 students from a first-year undergraduate Python course. Students attempted the programming task on pen and paper, and were encouraged to describe their problem-solving thought processes as pseudocode if they struggled to generate a Python-based solution. Figure \ref{fig:student_rsm} illustrates one example of a student's exploration, highlighting the opportunity for multiple solution pathways using representations that do not involve the relevant canonical method.

\subsection{Python Library}
We developed an open-source Python library\footnote{\url{https://github.com/augmented-human-lab/polar-ble-python}} as a software package that can be installed using Python's \textit{pip} package management system. 
We use a consumer-grade wearable heart-rate sensor from Polar 
to facilitate the use of sensor data in our implementation. The Polar Verity Sense\footnote{\url{https://www.polar.com/en/products/accessories/polar-verity-sense}} wristband is an optical heart rate sensor that captures physiological photoplethysmography (PPG) data, which can be used to measure features of Heart Rate Variability (HRV). We employ a Bluetooth Low Energy (BLE) communication protocol for wireless data transmission between the wearable sensor and our library. All operations for device connectivity and data processing are handled automatically by our library. For example, to access heart-rate data, users simply call a function named ``get\_latest\_heart\_rate()'' in their program. Thus, our library offers seamless plug-and-play use of sensor data, eliminating complicated technical configurations to simplify its use in a classroom context.

\section{Evaluation}

\subsection{Participants}
Our evaluation involved 20 students (9 male, 11 female) who were enrolled in an undergraduate CS1 Python programming course. At the time of our evaluation, students were in their third week of their semester and had not yet been introduced to lists. 

\subsection{Procedure}
\begin{figure}[t]
    \centering
    \includegraphics[width=\linewidth]{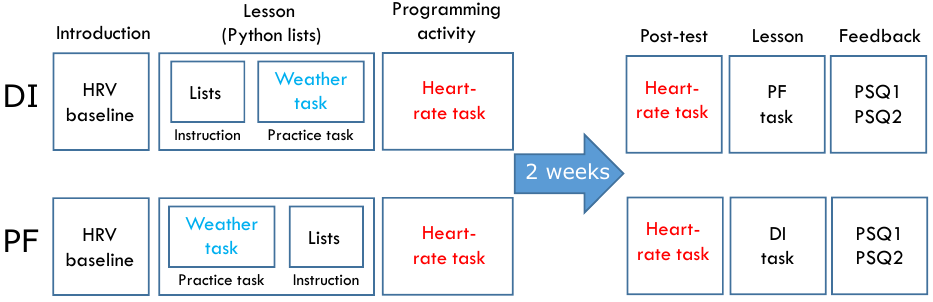}
    \caption{Overview of the study procedure for the evaluation of our PF learning experience (against DI as a control).}
    \label{fig:schematic}
    \vspace{-15pt}
\end{figure}

We designed  a controlled study with a subsequent short-term follow-up phase. Both the controlled study and the follow-up were conducted as individual sessions for each student and scheduled two weeks apart. Figure \ref{fig:schematic} provides an overview of the evaluation protocol we followed, which we expand on below.

\subsubsection{Initial Session (90 minutes)}
The initial session composed of a short introduction (15 minutes), a lesson on Python lists (45 minutes), and a programming activity (30 minutes). We randomly allocated half of the students to the experimental condition, where PF was used as the instructional strategy for the lesson on lists. The remaining students served as the control condition, learning about lists through traditional DI.

The introduction was used to brief students on the agenda of the session and to establish a baseline period of neutral cognitive load (see Section \ref{ch:ppg} for the calculation of cognitive load from PPG data). Students were asked to wear the heart-rate sensor on their forearm and relax for 5 minutes to collect the PPG data to determine this baseline value. The sensor was worn for the remainder of the study, enabling the use of heart-rate data in the programming tasks and the continued collection of PPG data to infer cognitive load.

The lesson on lists involved a 30-minute practice task and a 15-minute instructional unit. We developed the practice task for this lesson by creating an isomorphic version of our programming task on lists described in Section \ref{ch:3.1}. The isomorphic exercise was based on an identical sliding-window problem but relied on a temperature data stream from a weather station instead of live heart-rate data (and required tracking the most recent seven readings). In the experimental PF condition, students initially attempted the practice task, followed by a consolidation phase where shortcomings of their solutions were discussed and list concepts were explained. Conversely, the control DI condition began with an introduction to lists, after which students were given the same practice task. The session concluded with all students engaging in our programming task using heart-rate data where they applied their new knowledge of lists.

\subsubsection{Follow-up Session (45 minutes)}
Students returned two weeks later for their follow-up session, which included a post-test (10 minutes), a short lesson on a new concept (30 minutes), and a feedback session (5 minutes). We asked students to wear the sensor and engage in a post-test activity where they re-attempted the heart-rate-based programming task from their first session. Our focus in this segment was to gauge students' performance in delayed tasks after receiving instruction. 

Following the post-test, students participated in a short lesson on Python dictionaries, which was taught using the instructional condition they had not received in their initial session. Thus, students who experienced the PF condition in their first session received DI in their second session and vice versa. The motivation for this approach was to provide all students the opportunity to express their perceptions of the learning experiences provided by both DI and PF.
At the end of the lesson, we administered a questionnaire via Google Forms containing two open-ended questions: 
\begin{openquestions}
    \smallskip
    \item Please reflect on your experience with the study and share any thoughts you have about learning Python lists and dictionaries.
    \smallskip
    \item Each of the two sessions that you attended used a different strategy to teach a programming concept (list or dictionary). Which of the sessions worked better for you and why?
\end{openquestions}
These questions aimed to gather insights into students' perceptions and experiences of learning Python programming with DI and PF.

\subsection{Data Analysis}
We received completed questionnaires from 16 (out of 20) students who participated in our study and analysed their data.

\subsubsection{Programming performance}
The two programs (weather station and heart-rate) developed by each student during the initial session (Python lists) were used to explore how well they constructed solutions. We divided the collected data into two groups: control (DI; $n = 9$) and experimental (PF; $n = 7$). Unit testing was performed to determine if these programs produced the correct outputs for a given set of input data streams. In addition, we conducted a static analysis of the students' code to understand how their solution implementations compared to the canonical method and to classify any errors.

Similar measures were derived for the post-test programming task (heart-rate) in the follow-up session. This provided paired performance data on the heart-rate programming task for each student. 
We assessed learning outcomes by evaluating the changes in students' performance on this programming task that occurred over the two-week interval between sessions.

\subsubsection{Cognitive Load} \label{ch:ppg}
We established a data processing pipeline using HeartPy \cite{van2019heartpy} to infer cognitive load from the PPG data collected during the first session. PPG is a physiological signal representing blood volume pulse that can be used to compute Heart Rate Variability (HRV) by analysing the time intervals between successive peaks in its waveform. HRV is a reliable marker of cognitive load \cite{eija2010}, with a decrease in HRV indicating an increase in cognitive load \cite{thayer2009heart}. 
We calculate the root mean square of successive differences between normal heartbeats (RMSSD), which is a widely used time-domain HRV metric \cite{shaffer2017overview}, to extract cognitive load measures in our analysis. Changes in students' RMSSD values relative to their baseline at the start of the first session were used to infer the effects of DI and PF on cognitive load.

\subsubsection{Qualitative Analysis}
Thematic analysis was performed on the free-form responses to understand students' experiences and preferences of the learning strategies introduced. Following the guidelines by Braun and Clarke \cite{braun2006using}, we tagged responses to \textit{PSQ1} and \textit{PSQ2} and synthesised them into high-level themes.

\section{Results \& Discussion} \label{ch:7}
\subsection{Student Performance} \label{ch:7.1}
A summary of students' performance in their respective tasks on Python lists during the initial session and the follow-up session are presented in Table \ref{tab:results}.

\begin{table}[t]
\centering
\begin{adjustbox}{width=\columnwidth}
\begin{tabular}{|l|c|c|c|}
\hline
\multirow{2}{*}{\textbf{Group}} & \multicolumn{2}{c|}{\cellcolor{tableheadercolour}\textbf{Initial Success}} & \cellcolor{yellow!30}\textbf{Post-test Success} \\ \cline{2-4} 
& \textbf{Weather} & \textbf{Heart-Rate} & \textbf{Heart-Rate} \\ \hline
\textbf{DI ($n=9$)} & 8/9 & 8/9 & 6/9 \\ \hline
\textbf{PF ($n=7$)} & 2/7 & 7/7 & 7/7 \\ \hline
\end{tabular}
\end{adjustbox}
\caption{Summary of students' success in their initial tasks and the post-test task 2 weeks-later.}
\label{tab:results}
\vspace{-25pt}
\end{table}

\subsubsection{Learning with Direct Instruction} 
Students in the control group ($n = 9$) demonstrated high proficiency using lists during the practice task on weather station data, which they attempted immediately after receiving instruction.
Eight of the nine programs analysed executed successfully against our test cases and displayed optimal use of the list data structure.

When developing the solution for the weather station problem, students who received direct instruction adopted a concept-first approach, using their newly acquired knowledge of lists to guide the overall structure of their program. The example program shown in Listing \ref{listing:di_list} illustrates the standard solution procedure followed. Students used a conditional code block based on the size of the list to determine when data should be added or removed. In this instance, the student applied the \texttt{len} function to assess the length of the list, followed by the \texttt{append} and \texttt{pop} methods to add or remove values from the list, respectively. All accepted solutions were constructed using appropriate list methods taught during the learning phase. The \texttt{append} method was a popular choice for adding data to a list and was seen used across all programs. In contrast, we observed that a few students opted for alternatives to the `\texttt{pop}' method, such as \texttt{del}, \texttt{remove} and slicing, for deleting list items. Providing students with an initial conceptual understanding of lists prior to engaging in hands-on programming may have motivated a deeper exploration of list operations beyond those presented in our starter Python tutorial. Naturally, the results from the heart-rate-based programming task that followed shortly afterwards revealed identical performance, with the same eight students achieving correct solutions.

\begin{lstlisting}[language=Python, caption=Python program on Lists created by a student in the control (direct instruction) group, label=listing:di_list]
from weather_station import *

output_list = []
while NEW_DAY_AVAILABLE: # This simulates a new day
    temperature_today = get_daily_temperature()

    if len(output_list) < 7:
        output_list.append(temperature_today)
    else:
        output_list.pop(0)
        output_list.append(temperature_today)
        
    format_string = str(output_list)[1:-1]
    print(f"The latest 7 temperatures: {format_string}.")
\end{lstlisting}

\subsubsection{Learning with Productive Failure}
Requiring students in the experimental group ($n = 7$) to tackle the initial practice task on weather station data without adequate knowledge of lists elicited learning behaviours consistent with the expectations described by Kapur et al. \cite{kapur2012productive}. In the students' problem-solving efforts, there was evidence for the \textit{activation of prior knowledge}. This included initial ideas on the list concept and the generation of representations and solution methods using other programming concepts previously learned in class. Listing \ref{listing:pf_list}, for example, highlights an instance where the programming task prompted the recollection of knowledge about list-like linear data structures that the student may have acquired in the past. Although the student was ultimately unable to accomplish the task, their attempted solution suggests that PF could help reinforce and reactivate their understanding of concepts they may have previously encountered but not fully grasped. Indeed, students drawing on their relevant prior skills was a common approach observed across all seven programs, with techniques involving string handling, tuples and temporary variables employed in the solution generation process.

These behaviours also shed light on another key affordance of a PF learning strategy, which is the opportunity to explore \textit{multiple solution pathways} that may diverge from the intended canonical solution. This aligns with multiple conceptions theory proposed by Margulieux et al., in which the comparison of alternative solutions is a key factor for developing robust conceptual knowledge \cite{margulieux2021when}.
While no student succeeded in developing the optimal solution for the task, we identified two programs that achieved the correct functionality and satisfied all our test cases. One of these programs relied on string concatenation operations, such as \texttt{temperature = temperature + `, ' + temperature\_today}, to implement the same behaviour as the \texttt{append} list method. 

Static analysis of the five other solutions revealed that failure typically occurred in instances requiring knowledge of the canonical methodology. Common errors included incorrect logical statements involving variable scoping, data size checks, and data entry and removal. One such error can be seen in Line 13 of Listing \ref{listing:pf_list}, where the student tries to remove a data item by assigning its value to a secondary list. The errors we observed align with Kapur et al.'s guidelines for ``desirable failure''~\cite{kapur2012designing}, as students demonstrated an underlying conceptual awareness of the problem-solving task during their generation and exploration phase. This was particularly important for enabling an effective consolidation and knowledge assembly phase, in which students could reflect on their approach and develop a robust understanding of the canonical list-based solution. The outcomes of this learning process became evident in the subsequent heart-rate-based programming task, where all students optimally applied lists to construct their solutions.

\begin{lstlisting}[language=Python, caption=Python program on Lists created by a student in the experimental (productive failure) group, label=listing:pf_list]
from weather_station import *

temperature_list = []
discarded_list = []
temperature_count = 0
while NEW_DAY_AVAILABLE: # This simulates a new day
    temperature_today = get_daily_temperature()
	
    temperature_list += [temperature_today]
    temperature_count += 1
    
    if temperature_count > 7:
        discarded_list += [temperature_list[0]]	
        temperature_count = 0
	
    print(temperature_list)
\end{lstlisting}

\subsubsection{Learning Outcomes} \label{ch:7.1.3}
Our findings indicate that in scenarios incorporating both an instructional unit and a problem-solving task, students produced favourable outcomes regardless of the learning strategy employed. However, the results that emerged two weeks later, when students revisited the heart-rate programming task in the post-test, suggested that these effects may have been short-lived in the control group (DI). A total of three out of nine students failed to reproduce the expected solution for the task. Among them were two students who had previously performed successfully and one who had consistently failed since their first task two weeks earlier. The performance of this group raises the possibility that direct instruction is prone to ``unproductive success'', where the illusion of near transfer of knowledge may appear without authentic learning processes taking place. The views shared by students also echoed concerns about poor knowledge retention, stating they had ``\emph{easily forgotten despite learning it a mere 2 weeks ago}'' and that it ``\emph{was a struggle to remember the formalities of setting up a list}''.

All students in the experimental group (PF) showed greater persistence in maintaining their initial success on the heart-rate data problem when faced with the same task two weeks later. By grappling with novel problem-solving tasks before receiving instruction, students may be more likely to internalise concepts and apply them effectively in the future. 

\subsection{Student Physiology} \label{ch:7.2}
\begin{figure}[b]
    \vspace{-15pt}
    \centering
    \includegraphics[width=0.9\linewidth]{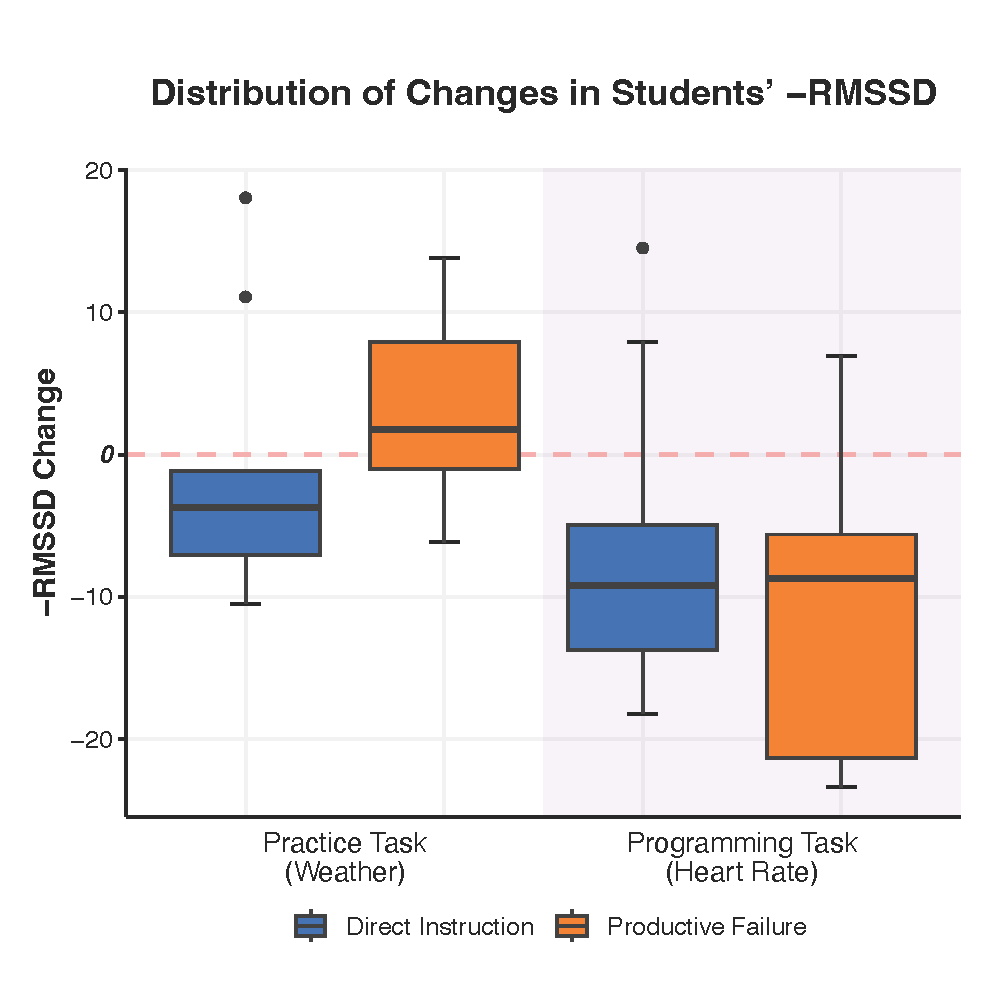}
    \caption{Distribution of changes in students' RMSSD during the practice (weather) task and the programming (heart-rate) task. A -RMSSD change that is less than 0 correlates to lowered cognitive load.}
    \label{fig:ch7_rmssd_plot}
\end{figure}

Figure \ref{fig:ch7_rmssd_plot} presents the average changes in the root mean square of successive differences between normal heartbeats (RMSSD) observed among students under different instructional conditions. Since RMSSD and cognitive load are negatively correlated, we illustrate the observed changes as $-RMSSD$ for better readability (an increase in $-RMSSD$ indicates increased cognitive load).

We observed a general trend of decreased cognitive load as students in both DI and PF groups progressed from the practice (weather) task to the programming (heart-rate) task.  Students in the experimental (PF) group appeared to experience an increased cognitive load during their practice task using the weather data. Since this task was introduced before instruction in the PF group, it likely demanded a greater mental effort.  On the heart-rate programming task, students in the PF and DI groups exhibited a similar change in their cognitive load overall, although students in the PF group had a much larger reduction in load when compared to their pre-instruction practice (weather) task.

Our findings mirror the core principles of Cognitive Load Theory \cite{sweller1994cognitive}, which emphasise the role of well-structured instruction in reducing cognitive load to facilitate successful learning. The lowered cognitive load induced by DI and PF in the respective tasks after instruction likely contributed to the learning outcomes discussed in Section \ref{ch:7.1.3}. Notably, students who received instruction after PF had the highest decline in cognitive load and consequently demonstrated the best overall learning performance. As hypothesised in Kapur et al.'s theory of Productive Failure, appropriately subjecting students to an initially heavy cognitive load may be fruitful for learning \cite{kapur2016examining}. 

\subsection{Student Perceptions} \label{ch:7.3}
We received a variety of responses to \textit{PSQ1} and \textit{PSQ2}, which asked students to reflect on their experiences learning Python and their preferred strategy between direct instruction and productive failure. Despite indications supporting more robust learning with PF, 
sentiment was more or less evenly split between the approaches, with 9 of the 16 students choosing PF as their favoured strategy.
Three central themes emerged regarding the considerations for choosing the preferred learning method, which were common across responses from all students. We use the terms \emph{effort}, \emph{feedback} and \emph{enjoyment} to describe these three themes. Although these themes were common to all students, their perspectives on learning effort, feedback, and enjoyment varied based on their choice of learning strategy, which we expand on below.

\subsubsection{Effort to Learn}
Students who preferred direct instruction commented on the efficiency of the approach. They appreciated being ``\emph{taught everything before I started work on the question}'', which they felt reduced their effort and helped them ``\emph{reach the objective faster}''. References to tags such as `fast', `quick', and `short-time' were popular among the students, appearing in six of the nine responses. Some students offered criticisms regarding the effort required to learn using productive failure, remarking that they ``\emph{doubt I would spot [errors] in a similar speed if I were to encounter this issue in the [PF] session}'' and that it would be ``\emph{challenging to solve the activities without knowing the direct option to solve it}''.

In contrast, those who preferred productive failure tended to welcome the effort needed to overcome learning challenges when expressing their thoughts about this strategy. Students viewed the difficulties encountered during the problem-solving process as beneficial learning experiences, stating that ``\emph{having to work through my memory and try and fail was a much more effective method, as it helped to ingrain the syntax and concepts in my mind better}''. The freedom to explore different ideas and solution representations encouraged students to ``\emph{experiment}'' and provided ``\emph{a starting point for [me] to think from and try to come up with a solution}''.

\subsubsection{Learning from Feedback}
Students preferring direct instruction reported that the role of instructors and educational resources in providing feedback were key factors for their choice of learning method. Their limited knowledge of Python at the time may have prompted them to frequently rely on the instructor's assistance during the learning phase. Representative comments include, ``\emph{[instructor\_name] provided us with information on [Python] dictionaries which I was then able to apply to efficiently}'' and ``\emph{I was able to get through questions pretty easily with the help of [instructor\_name] supporting me}''. 

Students who preferred productive failure valued the feedback they received from the mistakes made during their programming tasks. The ``\emph{firsthand experience [of] doing the programming first and slowly realising the mistake made during the coding}'' cultivated learning processes centred around students iteratively generating and reflecting on their failures. 

\subsubsection{Enjoyment of Learning}
All students appeared to enjoy the experience of both learning strategies introduced in the study. Those preferring direct instruction described their experience of learning Python programming as ``\emph{cool}'', ``\emph{great}'', and ``\emph{interesting}'' (6 comments). Since the concepts taught were a novelty to most students, there was much satisfaction in getting to create ``\emph{complete}'' and ``\emph{functional}'' programs that explored ``\emph{how [Python] lists could be used}''. 
The belief that they could succeed in their programming tasks inspired confidence among the students, which may have led to increased enjoyment. For example, when reflecting on why DI was a better learning strategy for them, one student notes that it ``\emph{improve[d] my confidence ... even though it was a short activity}''.

Students preferring productive failure found the highly engaging nature of their learning experience particularly enjoyable. The absence of immediate instructor intervention in this strategy allowed students to ``\emph{feel like I came up with the solutions myself}''. This sense of ownership over their solutions motivated students to continue exploring the concepts learnt beyond the classroom, as they felt they were ``\emph{more likely to know how to apply it in the future}''. Considering the relationship between motivation and learning, using a productive failure-based strategy could stimulate intrinsic motivation, making ``\emph{the session feel more rewarding}'' and ``\emph{memorable}''.

\section{Conclusion}
Productive failure is a learning approach that has gained popularity in STEM subjects, but its effectiveness in programming education remains under-explored. In this study, we compared Productive Failure (PF) with Direct Instruction (DI) for teaching novices about Python lists.  Our key findings revealed that while initial learning outcomes were similar between the PF and DI groups, students in the PF group exhibited better knowledge retention on a post-test task delayed by two weeks. Additionally, physiological data indicated a larger reduction in cognitive load for PF students when programming after they had received instruction.  
Although these results are promising, 
the nature of our data collection limited the number of participants, and future work should look to run controlled studies at a much larger scale.  In addition, 
collecting higher fidelity measures (such as with EEG) could provide clearer insights into factors such as cognitive load.
Future work should also explore new PF activities that target different programming concepts.